\documentstyle[]{mn}
\def\gs{\mathrel{\raise0.35ex\hbox{$\scriptstyle >$}\kern-0.6em 
les
\lower0.40ex\hbox{{$\scriptstyle \sim$}}}}
\def\ls{\mathrel{\raise0.35ex\hbox{$\scriptstyle <$}\kern-0.6em 
ggles
\lower0.40ex\hbox{{$\scriptstyle \sim$}}}}
\def\ltorder{
\mathrel{\raise.3ex\hbox{$<$}\mkern-14mu\lower0.6ex\hbox{$\sim$}}
}
\def\gtorder{
\mathrel{\raise.3ex\hbox{$>$}\mkern-14mu\lower0.6ex\hbox{$\sim$}}
}
\def\msun{\>{\rm M_{\odot}}}

\begin{document}

\title[Stellar contributors to the hard X-ray background?]
{Stellar contributors to the hard X-ray background?}
\author[Priyamvada Natarajan \& Omar Almaini]
{Priyamvada Natarajan$^{1,2}$ \& Omar Almaini$^3$\\
$^1$ Institute of Astronomy, Madingley Road, 
Cambridge CB3 0HA, U. K.\\ 
$^2$ Department of Astronomy, Yale University,
New Haven, CT 06520-8101, U. S. A.\\
$^3$ Institute for Astronomy, 
 University of Edinburgh, 
Royal Observatory, 
Blackford Hill, Edinburgh EH9 3HJ}

\label{firstpage}
\maketitle

\begin{abstract}
We use simple energetic arguments to estimate the contribution of
massive X-ray binaries and supernova remnants to the cosmic X-ray
background (XRB) at energies in excess of 2 keV. Recent surveys have
shown that AGN probably account for most of the hard XRB ($E\,>\,2$
keV), but there have been many suggestions that star-forming galaxies
could emerge at fainter fluxes and perhaps account for a significant
fraction of the soft and hard X-ray energy density. Assuming that the
formation rate of massive X-ray binaries (MXRBs) traces the global
star-formation rate, we find that their integrated contribution to the
hard XRB can be estimated and is shown to be small (at less than the
1\% level). Similarly, the integrated flux of SN is also shown to be
insignificant, or at most comparable to MXRBs. AGN therefore remain
the most viable candidates for producing the hard XRB, unless
additional processes can be shown to dominate the global hard X-ray
emission in distant starburst galaxies.
\end{abstract}

\section{Introduction}

Obscured AGN are now considered to be the most likely explanation for
the bulk of the hard XRB. If this hypothesis is correct, the total
energy output from these hidden AGN must exceed that of `ordinary'
(broad-line) objects by almost an order of magnitude (Fabian et al.
1998), with wide ranging implications. Existing deep X-ray surveys
have already revealed what could be the `tip of the iceberg' of this
population. Several unambiguous cases of obscured QSOs have been
detected (e.g. Almaini et al. 1995, Boyle et al.  1998, Fabian et al
2000) while at the faintest X-ray fluxes there is strong evidence for
a large population of X-ray luminous emission-line galaxies (Boyle et
al. 1995, Griffiths et al. 1996, McHardy et al. 1998).  The true
nature of this galaxy population remains controversial; however
statistical cross-correlation studies clearly show that objects with a
galaxy-like morphology contribute about $\sim 30 \% $ of the soft ($<
1$ keV) XRB (Roche et al. 1995, Almaini et al. 1997).  Results from
the ultra-deep {\it ROSAT} survey of Hasinger et al. (1998) suggest
that most of these `galaxies' contain AGN (Schmidt et al. 1998), but
the statistical sample is small.  In addition, many of these faint
sources were classified on the basis of high-ionisation optical
coronal lines, which could (in principle) arise from supernova
activity.  In summary, there is no doubt that most of the $0.5-2\,$keV
XRB is produced by AGN, but there is the realistic possibility of a
non-AGN contribution at the $10-20\%$ level.

Work has recently been undertaken at harder energies with {\it ASCA}
and {\it BeppoSAX}\, resolving $\sim 30\% $ of the harder, $2-10$\,keV
background. Many of these sources have now been identified, and most
appear to be absorbed AGN (Georgantopoulos et al. 1997, Fiore et al.
1999). Nevertheless, there is still the possibility of a starburst
contribution at fainter X-ray fluxes (Moran et al. 1999). Only the
deepest surveys with {\it Chandra} and {\it XMM} will resolve this
issue. The latest results from the first such survey probe an order of
magnitude deeper than before (Mushotzky et al. 2000) and claim to have
resolved $\sim$ 75\% of the hard XRB.  The nature of these Chandra
sources is so far unclear, but the most likely explanation seems to be
either absorbed AGN or possibly the first generation of quasars at
very high redshift.

In this letter, we attempt to quantify the potential contribution of
star-forming galaxies to the energy density of the XRB in a globally
averaged sense.  While this in itself is not a new suggestion (see
Griffiths \& Padovani 1990 and Treyer et al. 1992), recent
determinations of the integrated star-forming history of the Universe
(Madau et al. 1996; Lilly et al. 1996; Blain et al. 1999) enable a
quantitative estimate of the X-ray emission from stellar processes at
high redshift. Extrapolations based on no evolution models produce a
negligible contribution, however, star-formation rates and SN rates
were significantly higher during early epochs and so their integrated
contribution over redshift can now be calculated.  We use a recent
determination of the star-formation rate of the Universe that combines
optical and infra-red data (Blain et al. 1999) with observed
properties of MXRBs and SN to estimate their probable contribution to
the energy density of the XRB.  Possible additional sources of hard
X-ray emission are outlined in Section 4.

\section{Massive X-ray binaries}

MXRBs are produced in large numbers during bursts of star formation
and are expected to dominate the X-ray emission from starburst
galaxies (David, Jones \& Forman 1992). MXRBs are composed of a
compact object (a neutron star or a black hole) accreting matter from
a close O - B (Be) companion. Hard X-ray emission {\it E} $\sim$ 20
keV results from the mass transfer process. About 50 candidate MXRBs
have been detected in the Milky Way. They have thermal spectra with
characteristic $T > 15$ keV and $L_{\rm X} \sim 10^{37}$ erg s$^{-1}$
in the $E > 2$ keV band (van Paradijs \& McClintock 1995). They are
very short-lived, with life-times $t_{\rm MXRB} \sim 10^{5-6}$ yr,
much shorter than low mass X-ray binaries (LMXBs) which can last up to
a Gyr. The contributions of MXRBs and LMXBs to the X-ray flux of our
Galaxy are comparable. In the Milky Way, the majority of the MXRBs are
X-ray pulsars, for which the accreting object is a magnetic neutron
star. In low metallicity environments, such as the Magellanic clouds,
the X-ray emission from MXRBs is enhanced, and hence the contribution
from starbursts at high redshift (and low metallicity) could be much
more significant than is observed locally (van Paradijs \& McClintock
1995).

Several papers have addressed the issue of the contribution of
star-forming galaxies (Bookbinder et al. 1980; Griffiths \& Padovani
1990; Lahav et al. 1993; Treyer \& Lahav 1996) and MXRBs to the hard
XRB (Treyer et al. 1992). In all these treatments, the evolution of
the X-ray luminosity function of galaxies is extrapolated to high
redshifts in order to estimate the contribution to the XRB. Both
groups concluded that newly forming galaxies could contribute up to
15\% of the hard XRB, modulo the various uncertainties in the modeling
of the evolution of the galaxy luminosity functions, and the
conversion factor from the infra-red flux to X-ray flux.  With the
recent progress in observational determination of the globally
averaged star-formation rate as a function of redshift (Madau et
al. 1996; Lilly et al. 1996), the calculation of the total integrated
emission in the hard band from MXRBs in vigorously star-forming
galaxies is relatively straight-forward, if one assumes that the MXRB
rate instantaneously tracks the star-formation rate.

\subsection{Spectra and contribution to the hard XRB}

We commence with the assumption that the MXRB rate as a function of
redshift is proportional to the total star formation rate density
$\rho_{\rm MXRB}(z)\,\propto\, \rho_{\rm SFR}(z)$. This amounts to
assuming that for every solar mass of material that produces stars
(given an IMF) with observed luminosity density, some fraction of that
is processed into massive X-ray binaries. David, Jones \& Forman
(1992) find a tight relationship between the X-ray luminosity and the
star formation rate in starburst galaxies from the IRAS Bright Galaxy
sample, although there are recent claims that the scatter in the local
group may be much larger. This proportionality is calibrated locally
using fiducial values for the Milky Way: a star-formation rate of
$1\,{\rm M_{\odot}}$ yr$^{-1}$ and a total integrated MXRB luminosity
of $\sim\,2.2 \times 10^{38}$ erg s$^{-1}$ in the hard X-ray band ($E
> 2$ keV). This MXRB luminosity is based on the compilation of Dalton
\& Sarazin (1995), using the measured X-ray luminosities of the 8
known galactic MXRBs with peak X-ray luminosities in excess of
$10^{36}\,{\rm ergs^{-1}}$.  The fraction of the X-ray luminosity that
is emitted in the hard band is estimated with a typical spectral
energy distribution $f(E)$ for an MXRB that is well-fit by,
\begin{eqnarray}
f(E)\,&\propto&\,E^{a},\,\,\,E\,<\,E_c\\
f(E)\,&\propto&\,E^{b} exp (\frac{E_c - E}{E_f})\,\,\,{\rm otherwise}
\end{eqnarray}
$E_c = 19$ keV, $a = -0.217$ and $b = -2.67$ normalized at $E = 1$ keV
such that $f(E) = 1$ keV$^{-1}$ (Treyer et al. 1992). With this SED
almost 90\% of the X-ray emission is in the hard-band at $z = 0$. An
effective k-correction needs to be applied to take into account the
red-shifting of this SED out of the hard-band, which is done through
the function,
\begin{eqnarray}
f_{\rm hard}(z)\,=\,{\frac{\int^{\infty}_{3(1+z)\,{\rm keV}}\,f(E)\,dE}
{\int^{\infty}_{0.1\,{\rm keV}}\,f(E)\,dE}}.
\end{eqnarray} 
Note that since the bulk of the star formation occurs at $z\,<\,2$,
including the effect of redshift MXRBs do indeed have the required
spectral shape to explain the XRB at $E\,\geq\,3$ keV.
The integrated contribution from MXRBs to the hard XRB energy density
is therefore;
\begin{eqnarray}
\rho_{\rm MXRB}(z)\,&=&\,\int^{z}_{z_{\rm max}}f_{\rm hard}(z)
\xi_1\,({\frac{\rho_{\rm SFR}(z)}{1\,{\rm M_{\odot}}\,{\rm yr}^{-1}}})
\nonumber \\ &\,& \times (1+z)^{-7/2}\,{\it dz}\,\,{\rm erg\,Mpc^{-3}},
\end{eqnarray}
where $\xi_1$ is the locally calibrated constant for $\Omega = 1$ and
$H_0 = 50\,{\rm km\,s^{-1}\,Mpc^{-1}}$. To illustrate the uncertainty
in $\xi_1$, we write it explicitly in terms of the integrated MXRB
luminosity function $\Sigma_i\,n_i\,L_{X_i}$,
\begin{eqnarray}
\xi_1 = 6.31 \times 10^{45}\,(\frac{\Sigma_i\,n_i\,L_{X_i}}{2 \times
10^{38}\,{\rm erg s^{-1}}}).
\end{eqnarray}
One can then substitute the parametric form for $\rho_{\rm SFR}(z)$
that is a good fit to both the optical and the far-infrared data from
Blain et al. (1999), namely;
\begin{eqnarray}
\rho_{\rm SFR}(z)\,&=&\,0.009 \times {10^{0.6592z}}\,
{\rm \msun yr^{-1} Mpc^{-3}}\,\,z\,\leq\,2.5 \nonumber\\
\rho_{\rm SFR}(z)\,&=&\,0.4\,{\rm \msun yr^{-1} Mpc^{-3}}\,\,z\,>\,2.5.
\end{eqnarray}
Note that star-formation in this model-fit continues at a constant
rate out to high redshift. Another model that also fits current
observations (the Gaussian model from Blain et al. 1999) produces very
similar numbers.  Integrating from $z=5$ to $z = 0$, the contribution
of MXRBs was found to be $2.26\,\times\,10^{54}$ erg Mpc$^{-3}$. The
total energy density in the XRB is $2.63\,\times\,10^{57}\,$ erg
Mpc$^{-3}$, 80\% of which is in the hard band (Comastri 1998). MXRBs
therefore contribute negligibly - of the order say 0.2\%.  Note that
even if the IMF is biased toward high masses, increasing say, the
total number of MXRBs by a factor of 10 or if the X-ray luminosities
are higher in lower metallicity environments (van Paradijs \&
McClintock 1995), which is the case in high redshift starbursts or if
X-ray luminosity is higher in very compact starbursts, then MXRB
contribution to the hard XRB can be boosted to at most the few \%
level.

Our estimate differs from the results of previous work by Griffiths \&
Padovani (1990) and Treyer et al. (1992) due to reasons discussed
below. Both those works had to assume detailed models for the relation
between detected X-ray luminosity and either the optical luminosity or
60$\mu$m luminosity of star-forming galaxies and their evolution with
redshift.  Depending on their choice of assumed luminosity function,
both studies concluded that star-forming galaxies could contribute
anywhere from 4\% - 25\% or so. Therefore, we find consistency with
their most conservative model.  

\section{The contribution from supernovae}

In a typical galaxy, the total flux from the integrated number of SN
over a Hubble time is roughly comparable to the luminosity of the central
AGN, as we show below. The total energy output from SN can be 
estimated using the integrated SN rate from a given galaxy over a 
Hubble time with an assumed typical rate of 
$1/{\rm galaxy}/100\,{\rm yr} = 10^{-2}/{\rm galaxy}/{\rm yr}$,
\begin{eqnarray}
E_{\rm SN}\,&\sim&\,f_x \times \nu_{\rm SN} \times t_H \times \epsilon_{\rm SN}\nonumber \\ & = & 2 \times 10^{59} f_x\,\,{\rm erg}
\end{eqnarray}
where the typical energy output from a SN $\epsilon_{\rm SN}= 10^{51}
{\rm erg\,s^{-1}}$, $t_H = 10\,{\rm Gyr}$ and $f_x$ is the fraction of 
energy emitted in the X-rays by a typical SN. Note that this energy 
output is comparable to that from the central engine in AGN for 
$f_x \sim 10^{-4}$,
\begin{eqnarray}
E_{\rm AGN}\,&\sim&\,\epsilon f_{\rm AGN} M_{\rm bh} c^2 \nonumber \\
&=& 2 \times 10^{55}\,(\frac{\epsilon}{0.1})\,(\frac{M_{\rm bh}}{10^7 \msun}
)\,{\rm erg}.
\end{eqnarray}
 
Since AGNs are the primary discrete sources that contribute to
the XRB, SN are expected to contribute as well. However, since the
fraction of energy expelled in a supernova explosion that is emitted
in X-rays is small, their contribution to the hard XRB is expected to
be limited.  Locally it is observed that hard X-rays can be produced
by SN remnants via other processes, namely, by the presence of rare
Crab-like objects, synchrotron radiation from shocks due to the
collision of expanding shells and the inverse Compton scattering of
infra-red photons off radio-emitting electrons (as possibly seen in
M82).  Detailed treatment of these mechanisms however, is beyond the
scope of this work since we are averaging over the population as a
whole, but we discuss additional sources of X-ray emission further in
Section 4.

\subsection{X-ray flux from SN}

The star formation rate per unit volume quoted above is used to
estimate the supernova rate per unit volume, wherein the variations
from specific galaxy types are averaged over and the calibration is
performed to the Milky Way (Madau, Della Valle \& Panagia 1998).  Of
the typical energy output from a SN, $E_{\rm SN}= 10^{51} {\rm
erg\,}$, only a very small fraction ($f_x$) is emitted in hard
X-rays. SN, in general, are rather inefficient in emitting X-rays,
unless the local conditions under which they explode and expand
include the presence of strong magnetic fields, or are in the dense
interiors of pre-existing HII regions. The X-ray luminosities of
detected bright SN and compact SN remnants are observed to be roughly
$L_{\rm X} \sim 3 \times 10^{37}$ erg s$^{-1}$ (Williams \& Chu 1995),
and with estimated mean ages of $\sim 1000$ years,
\begin{eqnarray}
f_x \sim
10^{-3} \times ({\frac{t_{\rm rem}}{1000\,{\rm yr}}}) \times
({\frac{L_{X}}{3 \times 10^{37}\,{\rm erg s^{-1}}}}),
\end{eqnarray}
where $t_{\rm rem}$ is the average age of the remnant.  Note that here
`age' refers to the period that the remnant remains X-ray bright in
the hard band, the overall life-time is expected to be much longer, of
the order of $10^5$ years or so.  The
integrated contribution from SN to the total energy density of the
XRB is given by,
\begin{eqnarray}
\rho_{\rm SN}(z)\,&=&\,\int^{z}_{z_{\rm max}}\xi_2\,{E_{\rm SN}}\,{f_x}\,
{n_{\rm SN}(z)}\nonumber \\ 
&\,&(1+z)^{-7/2}\,{\rm dz}\,\,{\rm erg\,Mpc^{-3}},
\end{eqnarray}
where $f_x$ is the efficiency of hard X-ray emission by the SN,
$E_{\rm SN}$ is the typical energy output of a SN, $E_{\rm SN} \sim
10^{51} {\rm erg}$ and ${n_{\rm SN}(z)}$ is the supernova rate in
units of number per year per Mpc$^{-3}$.  Integrating the above
equation using ${n_{\rm SN}(z)}$ computed from the star-formation rate
in equation (5) for Type IIs and Type Ias, contributions of
$1.95\,\times\,10^{54}$ erg Mpc$^{-3}$ [Type II] and
$[5.89,6.03,3.63]\,\times\,10^{53}$ erg Mpc$^{-3}$ [Type Ias] (with
time delays of [0.3, 1, 3] Gyr assumed between the SN explosion and
the collapse of the primary star) were found. Clearly, Type IIs are
more important than Type Ias, but their contribution to the hard XRB
is only at the $0.1\% $ level. Therefore, X-ray emission from
ordinary SN cannot account for a sizeable fraction of the hard XRB;
however, it is worth noting that SN are believed to contribute
significantly to the gamma-ray background dominated by the
$\gamma$-ray line emission from the decay of $^{56}$Ni, primarily from
Type Ia's (see Watanabe et al. 1998).

\subsection{Supernovae in compact, dense environments}

SN exploding in high-density environments can rapidly reach very high
X-ray luminosities, perhaps high enough to power the broad-line region
of some AGN (e.g. Terlevich et al 1992). A good example is SN 1988Z,
detected by Fabian \& Terlevich (1996), the most distant SN yet
detected in X-rays.  These SN achieve maximum luminosity and very high
temperatures ($\sim 30$keV) soon after the explosion ($t\sim 1$ yr).
Thereafter they cool and the luminosity decays roughly as $L_x \propto
t^{-11/7}$ (see Fabian \& Terlevich 1992 and references therein) with
luminous lifetimes of $10-100$ years.

These events are much more efficient generators of X-ray
emission. Taking SN 1988Z as a prototypical example, one finds that
$\sim 2$ per cent of the total energy of the supernovae can be
eventually liberated in X-rays, an order of magnitude higher than
``ordinary'' supernovae. Such events will clearly boost the potential
supernovae contribution to the hard XRB. To place an upper bound on
this contribution, one can assume that {\em all} supernovae occur in
such environments and scale to the supernova rate as above. This leads
to an upper limit of $1-2 \% $ to the total XRB energy density.

\section{Additional sources of hard X-ray emission?}

In this work we concentrate on the `classic' sources of X-ray emission
from starburst galaxies, namely X-ray binaries and emission from
supernovae. However there are additional sources of X-ray emission
which, potentially, could significantly boost the starburst
contribution to the hard X-ray background. Taking these in turn:

\subsection{Compton scattering of IR photons}

There is plausible evidence in at least two local galaxies (M82 and
NGC3256) that IR photons can be Comptonised to X-ray energies by the
relativistic electrons produced by supernovae (Moran \& Lehnert 1997,
Moran, Lehnert \& Helfand 1999). Given the vast reservoir of IR
photons produced by starburst activity, even a relatively low
Comptonising efficiency can produce significant X-ray
emission. Estimating the expected X-ray luminosity is highly uncertain
however, and requires detailed knowledge of the geometry and energy
budget in the Synchrotron plasma. Nevertheless, such models are strong
candidates for explaining the hard X-ray emission in M82 and
NGC3256. The possibility that these processes are globally important
for starburst galaxies cannot be ruled out.

\subsection{Crab-like Synchrotron sources}

Crab-like SNRs (``plerions''; Weiler \& Panagia 1978) are rare
Synchrotron dominated X-ray sources. Synchrotron emission is unusual
in X-ray astronomy due to the exceptionally high magnetic fields
required and the short radiative lifetimes, but the handful of
Crab-like objects known are important, highly luminous exceptions. One
cannot exclude the possibility that extreme, Crab-like objects may be
more important in early star forming galaxies, although so far there is
no evidence to support this.

\subsection{Starburst driven super-winds}

Many local starburst galaxies show evidence for energetic outflowing
winds (e.g.  Heckman et al 1996). These winds can account for roughly
50 per cent of the X-ray luminosity in the soft (ROSAT) X-ray band but
with typical thermal temperatures of $<1$ keV they are not expected to
contribute significantly at higher energies.

\section{Conclusions and discussion}

In this letter, the contribution of MXRBs and SN to the hard XRB have
been computed using recent estimates of the globally averaged
star-formation rate in the Universe. It is found that MXRBs contribute
at the 1\% level and that the contribution of SN is probably
significantly lower. Taking into account the uncertainties in the
efficiency of hard X-ray emission and its unknown dependence on
metallicity, MXRBs can at most account for a few percent of the hard
XRB, despite their optimal spectral shape.  Likewise, if most early
supernovae occur in highly compact environments, their potential
contribution could rise significantly, but still produce only $\sim
1-2 \%$ of the hard XRB.  We conclude that AGN will dominate the
source counts in forthcoming X-ray surveys with {\rm Chandra} and {\rm
XMM} {\em unless} the X-ray emission from distant starburst galaxies
is dominated by alternative processes. Compton-scattering of IR
photons by relativistic electrons and/or a hitherto unrecognized
abundance of Crab-like synchrotron sources would boost the hard X-ray
emission from starburst galaxies significantly, but as yet the
universal role of such phenomena is unclear.

\section*{Acknowledgements}

Useful discussions with Philip Armitage, Guenther Hasinger and Piero
Madau are gratefully acknowledged.  PN thanks Andrew Blain for careful
and detailed comments on the manuscript. We also thank the referee,
David Helfand, for very constructive comments which tightened up the paper
significantly.

\end{document}